\documentstyle[aps,aps12,prb]{revtex}

\begin{document}

\title{Superfluidity of indirect excitons in a quantum dot.}
\author{Yu.E. Lozovik\cite{e-mail}$^{,a}$, S.A. Verzakov$^a$,
and M. Willander$^b$
}
\address{$^a$Institute of Spectroscopy, Russian Academy of Sciences,
142092 Troitsk, Moscow region, Russia}
\address{$^b$G\"oteborg University,
Chalmers University of Technology, S-41296, G\"oteborg, Sweden}

\maketitle

\begin{abstract}

The superfluidity and Bose-Einstein condensation of indirect excitons in
two-dimensional quantum dot are studied by path-integral Monte Carlo
simulations. The temperature dependence of superfluid and Bose-condensed
fraction are calculated at different strengths of interaction.
Using the Kosterlitz-Thouless recursion relations, we also predict
behavior of superfluid fraction in macroscopic large systems.

\end{abstract}

\draft

\section{Introduction}\label{s_Introduction}

Low-dimensional electron-hole systems in semiconductor heterostructures,
quantum dots and quantum wires are of increasing experimental and theoretical
attention now. Conceptually new physical effects can take place in these
systems.\cite{FQEH1,FQEH2} One of the most interesting examples of systems
considered is the structure that consists of two vertically connected
two-dimensional quantum wells. Under the action of laser radiation, each well
is occupied by carriers of different types, which form a two-layered
electron-hole system. As has been shown in Ref. \onlinecite{Lozovik_etal},
superfluidity of indirect excitons, which are bounded states of pairs of
spatially-separated electrons and holes, can appear in the system and
manifests itself as persistent electric currents in each two-dimensional
layer. The behavior of phase diagram in strong magnetic
field\cite{Lozovik_Berman} and Josephson effects\cite{Lozovik_Poushnov} have
been also studied. A number of interesting experiments for magnetoexcitons
with spatially-separated carriers in coupled quantum wells have been recently
carried out.\cite{Zrenner,Bayer1,Sivan}

The surface roughness can lead to the excitons localization\cite{Gevorkyan}
and thus it behaves like a natural quantum dot.\cite{Zrenner} Excitons can
also be confined in vertically coupled artificial quantum
dots.\cite{Kaputkina} Hence, the investigation of "lakes" of excitons is
essential.

The goal of this article is to study the onset of the superfluidity of
indirect excitons in two-dimensional (2D) quantum dot. In a 2D system at low
densities $n \ll a^{-2}$ ($a$ is exciton radius), potential barrier that is
generated by dipole-dipole repulsion of indirect excitons suppress exchanges
of constituent particles, electrons and holes.\cite{Lozovik_Berman}
Therefore the system of 2D indirect excitons at low density can be treated as
Bose system. (Other situations in which excitons can be 
assumed to be bosons were discussed in 
Refs. \onlinecite{Keldysh_Kozlov,Kallin,Vignale}.)
Moreover, under the same conditions indirect excitons can be
thought of as point-like particles with unidirected dipole momenta. Another
system which can be mapped onto this model is that of trapped Bose atoms
with dipole momenta induced by d.c. electric field or
standing electromagnetic wave in resonance with atomic transition. For the
sake of definiteness we present explanation below in terms of excitonic
cluster. In the absence of interaction between particles, we use a
semi-analytical method based on a recursion relation for canonical partition
functions. Clusters of interacting excitons are explored by the path-integral
Monte Carlo technique.

The paper is organized as follows. In the next section the model and
controlling dimensionless parameters are introduced. In Sec. \ref{s_Methods}
we outline the Monte Carlo technique and a semi-analytical approach.
Sec. \ref{s_Discussion} is devoted to the discussion of the results.

\section{The model Hamiltonian}\label{s_Hamiltonian}

As discussed above, we treat indirect exciton as a structureless boson with
effective mass ${m}$ and effective dipole moment {\bf d} perpendicular to the
plane of 2D quantum dot. The confining potential of the quantum dot we take
in a harmonic form. Hence, the model Hamiltonian is given by
\begin{eqnarray}
\label{eq_Hamiltonian}
\hat{H} =
\sum_{i=1}^{N}
\left[
-
\frac{\hbar^2}{2 m}
\Delta_i
+
\frac{m\omega^2|{\bf r}_i^2|}{2}
\right]
+
\sum_{i<j} \frac{d^2}{|{\bf r}_{i,j}|^3}.
\end{eqnarray}
Choosing as length unit ${r_0 \equiv \sqrt{\hbar/m\omega}}$, we introduce
dimensionless controlling parameters for the system at the finite temperature
${\beta^{-1}}$: dimensionless temperature
${T \equiv \beta^{-1}/(\hbar^2/(2 m r_0^2))}$ and interaction constant
${k \equiv (|{\bf d}|^2/r_0^3)/(\hbar^2/(2 m r_0^2))}$.

\section{Simulation methods and calculated properties}\label{s_Methods}

The canonical ensemble of noninteracting Bose system (${k=0}$) can be studied
with the help of simple semi-analytical procedure analogous to that has been
used for particles in a harmonic trap interacted by a harmonic
law.\cite{Brosens} The method that has been used is based on universal
connections between partition functions of canonical and grand canonical
ensembles
\begin{eqnarray}
\nonumber
Z(u) = \sum_N  u^N  Z_N,
\\
\nonumber
u \equiv exp(\beta\mu),
\\
\label{eq_Z_N_via_Z_u}
Z_N = \left.\frac{1}{N!}\frac{d^N}{du^N}Z(u)\right|_{u=0}.
\end{eqnarray}
Here $\mu$ is the chemical potential of the grand canonical ensemble and N
is the number of particles in the canonical ensembles. Starting from
Eq. (\ref{eq_Z_N_via_Z_u}), one can derive a recursion relation for the
canonical partition functions of a noninteracting Bose gas with an arbitrary
one-particle spectrum ${\varepsilon_i}$:\cite{Fermi_expr}
\begin{eqnarray}
\label{eq_Z_N_recursion}
Z_N(\beta)
=
\frac{1}{N}\sum_{n=0}^{N-1}
Z_n(\beta)Z_1((N-n)\beta),
\\
\nonumber
Z_0(\beta) = 1,\;\;\;\;\;Z_1(\beta) = \sum_{i}\exp(-\beta\varepsilon_i).
\end{eqnarray}
Hence, the numerical procedure consists in calculations of the canonical
one-particle partition functions at temperatures ${T/N}$, ${2T/N}$, ... ${T}$
and iterations of Eq. (\ref{eq_Z_N_recursion}). Generalization of the
Eq. (\ref{eq_Z_N_recursion}) to the case of partition function derivatives,
in form of which interesting thermodynamic averages (see this section below)
can be presented, is obvious.

Cluster of interacting excitons has been studied by the path-integral Monte
Carlo calculations (for comprehensive review of application the path-integral
method to the Bose systems see Ref. \onlinecite{Ceperley_RMP}). This
technique is based on the Trotter procedure as a result of which
thermodynamic averages on ${D}$ dimensional quantum system are expressed via
averages on ${D+1}$ dimensional classical system, which is created by
multiplication of the configuration space of the initial system along
imaginary time direction. Calculated in so called primitive approximation,
the potential energy of the effective classical system that corresponds to
the quantum model with Hamiltonian (\ref{eq_Hamiltonian}) is
\begin{eqnarray*}
\beta V_{eff} =
\frac{1}{PT}
\sum_{p=0}^{P-1}
\left(
\sum_{i=1}^{N}
\left[
\frac{|{\bf r}_i^{\;(p+1)} - {\bf r}_i^{\;(p)}|^2}{4/(PT)^2}
+
|{\bf r}_{i}^{(p)}|^2
\right]
+
\sum_{i<j}
\frac{k}{|{\bf r}_{i,j}^{(p)}|^3}
\right),
\end{eqnarray*}
where ${P}$ is the number of effective system layers, ${N}$ is the number
of particles in one layer, which is equal to the number of particles in
the original quantum system and ${{\bf r}_i^{\;(p)}}$ stands for the 2D
radius vector of ${i}$-th particle in the layer ${p}$. Bose statistics is
taken into account by summation over particles permutations $\cal P$ during
Monte Carlo simulation.\cite{Ceperley_RMP} The energy of the effective system
depends on $\cal P$ by assumption
${{\bf r}_i^{\;(P)}={\bf r}_{{\cal P}(i)}^{\;(0)}}$.

Our attention has been focused on superfluid properties of excitons. One of
the most dramatic manifestations of the superfluidity is the deviation of the
effective moment of inertia from its rigid or "classical" value
${I_{cl} \equiv \left< \sum_{i=1}^N m r_i^2\right>}$. By definition, the
effective moment of inertia is the linear response function to a uniform
rotation field with angular frequency ${\Omega}$
\begin{eqnarray*}
I_{eff} \equiv
\left.
\frac
{\partial}
{\partial \Omega}
\frac
{tr\{L_z \exp(-\beta (H-\Omega L_z))\}}
{tr\{\exp(-\beta (H-\Omega L_z))\}}
\right|_{\Omega = 0}
= \beta\left<L_z^2\right>.
\end{eqnarray*}
Only the normal fluid gives rise to the ${I_{eff}}$.\cite{Baym} Thus, the
superfluid fraction of the cluster is
\begin{eqnarray*}
\nu_s \equiv \frac{n_s}{n} = \frac{I_{cl} - I_{eff}}{I_{cl}}.
\end{eqnarray*}

In order to calculate ${\nu_s}$ in case of ideal gas in a trap
(i.e. at ${k=0}$, see definition of ${k}$ near Eq. (\ref{eq_Hamiltonian}))
by using above mentioned recurrent procedure, firstly, one should express
${I_{eff}}$ and ${I_{cl}}$ in form of partition function derivatives
\begin{eqnarray*}
I_{eff}
\equiv
\beta\left<\hat{L}_z^2\right>
=
\frac{1}{\beta Z}
\frac{\partial^2}{\partial\Omega^2}
\left.
Z^{\Omega}
\right|_{\Omega=0},
\\
I_{cl}
\equiv
\left<\sum_i m r^2_{i}\right>
=
\frac{1}{\beta\omega Z}
\frac{\partial}{\partial\omega}
Z.
\end{eqnarray*}
Here
${Z^{\Omega}\equiv tr\left\{\exp(-\beta(\hat{H}-\Omega\hat{L}_z))\right\}}$
is the partition function of the system rotating with the frequency
${\Omega}$, ${Z \equiv Z^{\Omega=0}}$ is the partition function of the system
at rest. Secondly, explicit expression for one-particle partition function
${Z^{\Omega}}$ is needed. The spectrum of the Hamiltonian
${\hat{H}^{\Omega} = \hat{H} - \Omega\hat{L}_z}$ at ${k=0}$  is
${\varepsilon_{n,m} = \hbar\omega(2n + |m| + 1) - \hbar\Omega n}$, where
${n = 0,1, \ldots}$ and ${m = 0,\pm 1, \pm 2, \ldots}$ are radial and angular
quantum numbers (${\hbar m}$ is the eigenvalue of the operator
${\hat{L}_z}$). After some algebra, one can write
\begin{eqnarray*}
Z_1^{\Omega}(\beta) =
[4
\sinh(\hbar\beta(\omega-\Omega)/2)\;
\sinh(\hbar\beta(\omega+\Omega)/2)
]^{-1}
\end{eqnarray*}

Calculation of superfluid fraction in case of $k \neq 0$ have been performed
by path-integral Monte Carlo simulations. Applying the Trotter procedure to
the quantum-statistics averages by which ${I_{cl}}$ and ${I_{eff}}$ are
defined, one takes\cite{Ceperley_RMP}
\begin{eqnarray*}
I_{eff} =
\frac{1}{P}
\left<
\sum_{i=1}^{N}\sum_{p=0}^{P-1} {\bf r}_i^{\;(p)} \cdot {\bf r}_i^{\;(p+1)}
\right>_P - 2T \left<|{\bf A}|^2\right>_P
\end{eqnarray*}
\begin{eqnarray*}
{\bf A} =
\frac{1}{2}
\sum_{i=1}^{N}\sum_{p=0}^{P-1} {\bf r}_i^{\;(p)} \times {\bf r}_i^{\;(p+1)}
\end{eqnarray*}

\begin{eqnarray*}
I_{cl} =
\frac{1}{P}
\left<
\sum_{i=1}^{N}\sum_{p=0}^{P-1} {\bf r}_i^{\;(p)} \cdot {\bf r}_i^{\;(p)}
\right>_P
\end{eqnarray*}
Here averagings are performed on the effective classical system, the moments
${I_{cl}}$ and ${I_{eff}}$ are measured in units of ${mr_0^2=\hbar/\omega}$.

The fraction of Bose-condensed particles we calculate by the method
introduced in Ref. \onlinecite{Krauth}: the maximum length of permutation
cycle in path-integral Monte Carlo simulation which has a nonzero probability
is the number of particles in the condensate.

\section{Discussion of the simulation results}\label{s_Discussion}

Calculated temperature dependencies of superfluid fraction are shown in
Fig. 1. One can see superfluidity suppression induced by interaction at a
fixed dot angular frequency. Condensate depletion is obvious too
(see Fig. 2). However after defining "average density"
${n \equiv N/(2 \pi <r^2>)}$ and plotting ${\nu_s}$  {\it vs.}
dimensionless temperature
${\alpha \equiv m/(\beta\hbar^2 n) = 2\pi I_{cl}/(\beta\hbar^2N^2)}$,
i.e. at a {\it fixed density}, one can see (Fig. 3) some rising of the
superfluid fraction with interaction. But interaction, of course, depletes
condensate even at fixed density (Fig. 4).

The investigated system is fully analogous to the atoms in a plane harmonic
trap. BEC occurs for noninteracting system in two-dimensional harmonic trap
in thermodynamic limit (contrary to the situation in homogeneous systems).
For systems with repulsive interaction there is only some critical
temperatures of transition to the new state without condensate
(see Ref. \onlinecite{Mullin} and references therein).

Assuming Kosterlitz-Thouless (KT) scenario, we have estimated the critical
temperature of the phase transition. Nonanalytic behavior of the free energy
of the system and consequently a phase transition can be observed only in
macroscopic systems. But in a computer simulation one deals with finite
systems. Hence, the extrapolating procedure is necessary. We define
thermodynamic limit as ${\omega\to 0}$, ${N\to\infty}$ at
${n\equiv N/(2 \pi <r^2>)=constant}$. In case of weak interaction this
definition coincides with requirement ${N\omega^2 = constant}$.

The extrapolation procedure (see Ref. \onlinecite{Ceperley_2D}) is based on
mapping of a 2D system of vortices to the 2D logarithmic gas, which has KT
transition.\cite{KT,Minnhagen}

The temperature of this gas, which is expressed in units of square of
logarithmic gas charge, is
${T_{cg} = m/(2\pi\hbar^2 n \beta) = \alpha/(2\pi)}$, the chemical
potential (in the same units) is ${\mu = - T_{cg} \beta E_c}$, where ${E_c}$
is the vortex core energy in superfluid, and dielectric constant is
${\varepsilon = 1/\nu_s}$. Kosterlitz-Thouless recursion relations lead to
the universal jump of ${1/(T_cg \varepsilon)}$ from ${4}$ to ${0}$ at the
transition temperature. But in the finite systems superfluid vanishes
smoothly enough (Figs. 1, 3). Such type of behavior can be analytically
accounted by integrating recursion relations up to size of system (not to
infinity).\cite{Ceperley_RMP}

The parameters of curve ${\nu_s(T_{cg})}$ are vortex core energy and
diameter, which are defined by fitting analytic approximation to the Monte
Carlo results. Then full integration of Kosterlitz-Thouless relations at
fixed vortex core energy and core diameter gives critical value of
temperature.

Scaling results for different values of interaction constant ${k}$ gives,
e.g., ${\alpha_{KT}(0.1) = 0.92}$, ${\alpha_{KT}(1) = 1.19}$.

\section*{Acknowledgments}
We wish to thank A.I.~Belousov for fruitful discussion. The work has been
supported by a grant from The Royal Swedish Academy of Science and INTAS.



\newpage
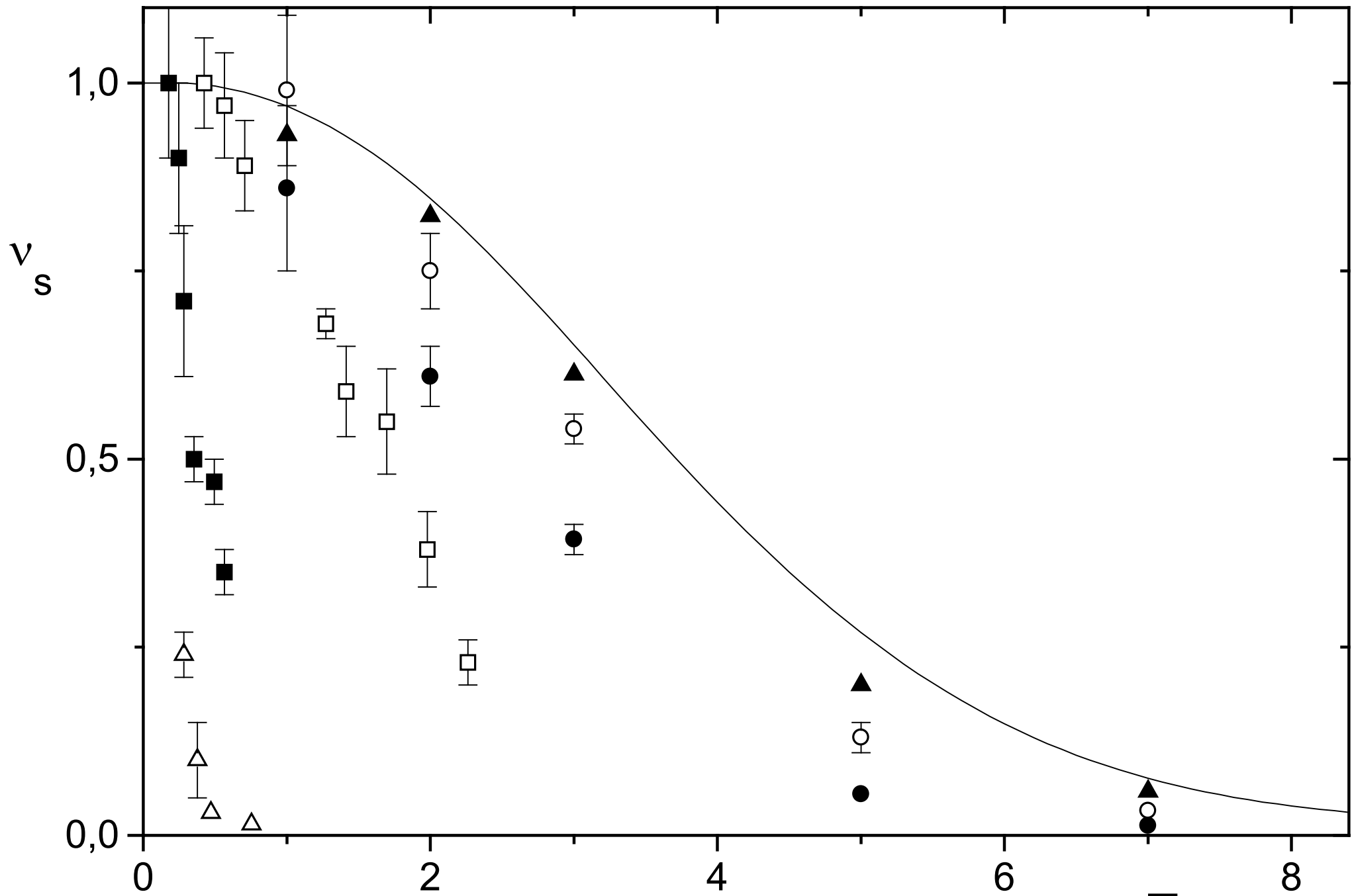
FIG. 1. Superfluid fraction $\nu_s$ {\it vs.} dimensionless temperature
$T=2/(\beta\hbar\omega)$ for excitonic cluster with
number of particles $N=37$:
solid line -- $k=0$ (noninteracting case), solid triangles -- $k=0.01$,
open circles -- $k=0.1$, solid circles -- $k=1$, open squares $k=2.38$,
solid squares $k=23.5$, open triangle $k=48.25$.

FIG. 2. Condensed fraction $\nu_0$ {\it vs.} $T=2/(\beta\hbar\omega)$
for excitonic cluster with number of particles $N=37$:
solid line -- $k=0$ (noninteracting case), open circles -- $k=0.1$,
solid circles -- $k=1$.

FIG. 3. Superfluid fraction $\nu_s$ {\it vs.}
$\alpha = 2\pi I_{cl}/(\beta\hbar^2N^2$:
solid line -- $k=0$ (noninteracting case), solid triangles -- $k=0.01$,
open circles -- $k=0.1$, solid circles -- $k=1$,
open squares -- $k=2.38$, solid squares -- $k=23.5$.

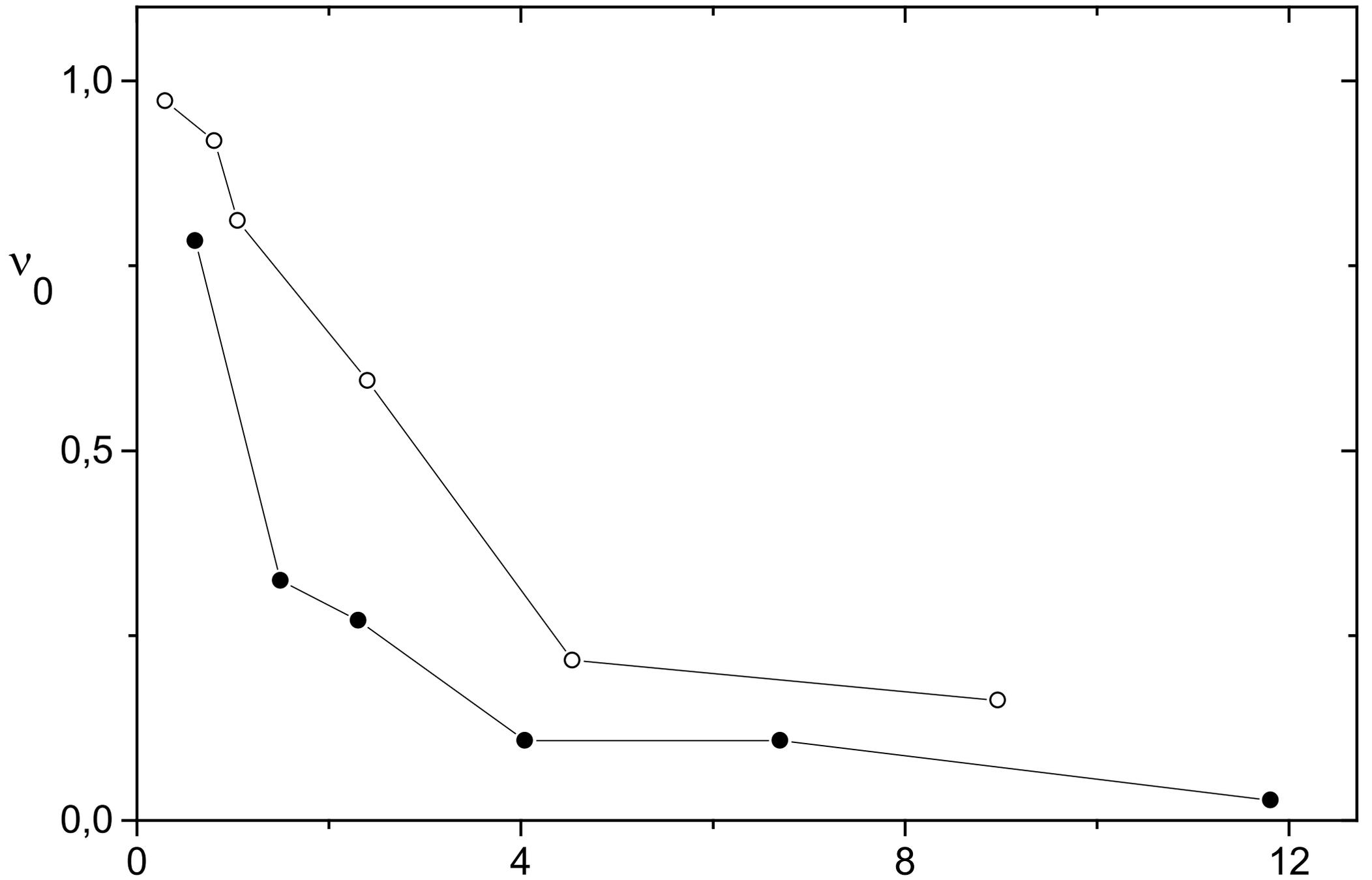
FIG. 4. Condensed fraction $\nu_0$ {\it vs.}
$\alpha = 2\pi I_{cl}/(\beta\hbar^2N^2$:
solid line -- $k=0$ (noninteracting case), open circles -- $k=0.1$,
solid circles -- $k=1$.


\begin{references}
\bibitem{Fermi_expr}
                    In case of Fermi statistics relation is
                    ${Z_N(\beta)=\frac{1}{N}\sum_{n=0}^{N-1}%
                    (-1)^{N-1-n}Z_n(\beta)Z_1((N-n)\beta)}$
\end{references}
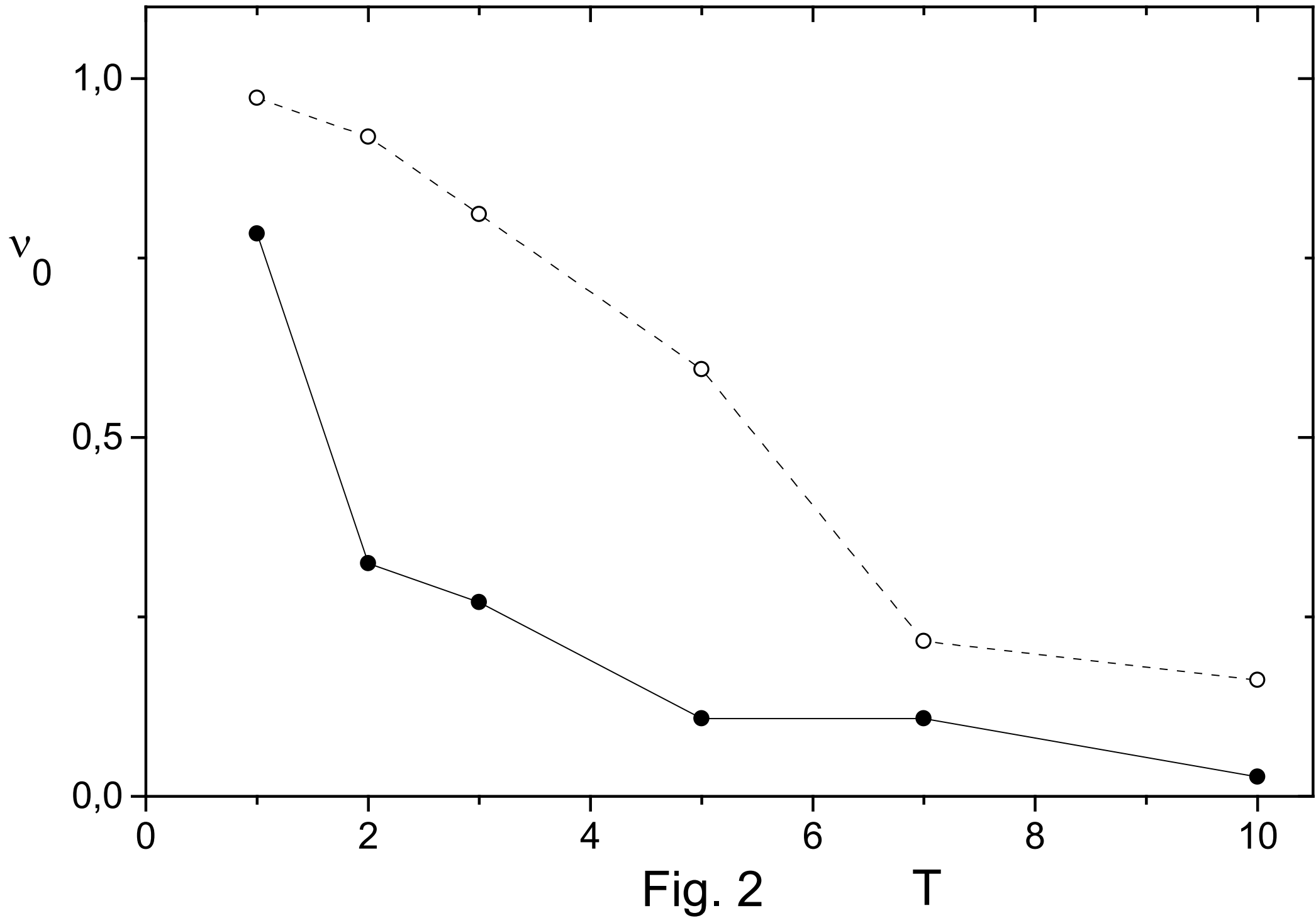
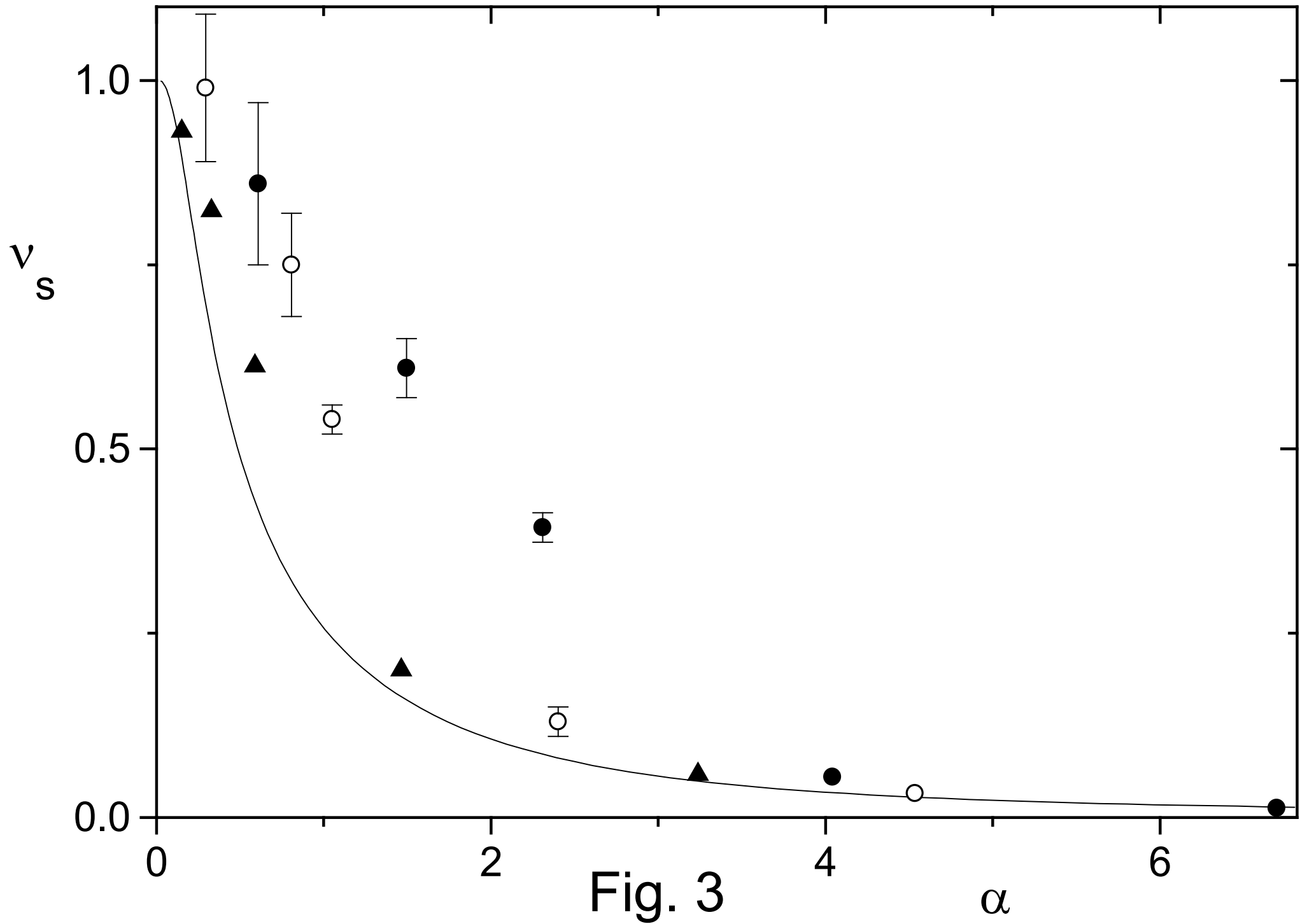
\end{document}